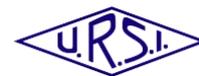

# FAST Observation and Results for Core Collapse Globular Cluster M15 and NGC 6517

Yu-Xiao Wu[(1)], De-Jiang Yin[(2)], Yu Pan[(1)], Li-Yun Zhang[(2)], Zhichen Pan[(3)(4)(5)], Lei Qian[(3)(4)(5)], Bao-Da Li[(2)], Yin-Feng Dai[(2)], Yao-Wei Li[(2)], Xing-Nan Zhang[(7)], Ming-Hui Li[(7)], Yi-Feng Li[(6)]

(1) School of Science, Chongqing University of Posts and Telecommunications, Chongqing 400000, China; e-mail: 2017213252@stu.cqupt.edu.cn
(2) College of Physics, Guizhou University, Guiyang 550025, China; e-mail: gs.djyin21@gzu.edu.cn
(3) National Astronomical Observatories, Chinese Academy of Science, 20A Datun Raod, Chaoyang District, Beijing 100101, China
(4) Guizhou Radio Astronomical Observatory, Guizhou University Guiyang 550000, China
(5) College of Astronomy and Space Sciences, University of Chinese Academy of Sciences, Chinese Academy of Sciences, Beijing 100101, China
(6) National Time Service Center, Chinese Academy of Sciences, Xi'an 710600, China
(7) State Key Laboratory of Public Big Data, Guizhou University, Guiyang 550025, China

## Abstract

Radio astronomy is part of radio science that developed rapidly in recent decades. In the research of radio astronomy, pulsars have always been an enduring popular research target. To find and observe more pulsars, large radio telescopes have been built all over the world. In this paper, we present our studies on pulsars in M15 and NGC 6517 with FAST, including monitoring pulsars in M15 and new pulsar discoveries in NGC 6517. All the previously known pulsars in M15 were detected without no new discoveries. Among them, M15C was still detectable by FAST, while it is assumed to fade out due to precession [1]. In NGC 6517, new pulsars were continues to be discovered and all of them are tend to be isolated pulsars. Currently, the number of pulsars in NGC 6517 is 17, much more than the predicted before [2].

## 1 Introduction

Since pulsars were initially observed and discovered[3], they had become one of the most popular astronomical research targets. Due to the significant properties of pulsars, such as extremely stable periods, high-intensity magnetic fields, and other characteristics, finding more pulsars broadens the view of physics in extreme conditions. Thus, a considerable number of radio telescopes were built, including the Five-hundred-meter Aperture Spherical radio Telescope (FAST [4]) which made surprising results recently.

FAST has the most enormous collecting area for radio waves among all the current single dish radio telescopes, with an illumination diameter of 300 m and an aperture efficiency of ~60%. Due to the remarkable sensitivity, surveys with FAST resulted in more than 600 new pulsar discovered [5, 6].

Globular clusters (GCs) are always one of the wish choices for finding exotic pulsars. Until the beginning of February 2023, 278 pulsars in total have been discovered in 38 GCs[1], and most of these pulsars are millisecond pulsars and/or binaries. The fastest spinning pulsar [7], pulsar neutron star systems such as M15C [8], and more than ten redbacks and black widows were found in GCs. Among the GCs, NGC 6517 and M15 contain a big number of pulsars and they are both core-collapsed clusters. There are 9 pulsars in M15, including 7 isolated, 1 binary with a neutron star companion, and 1 candidate that can be a binary[8, 9, 10]. NGC 6517 contains 17 pulsars [2, 10, 11].

In this paper, we summary the previous GC pulsar discoveries with FAST (Section 2), report our studies to M15 and NGC 6157. Section 3 and 4 are for M15 and NGC 6517, respectively. Section 5 is for summary.

## 2 GC Pulsar Search with FAST

The first new pulsar discovered by FAST was J1859-0131 with a period of 1.82 seconds [12]. It opened the first page of FAST's detection of Galactic plane pulsars and GC pulsars. Figure 1 are the summary for the GC pulsars discovered in recent years and they are sorted by telescopes. No Doubt that FAST and MeerKAT discovered most of the GC pulsars.

With eleven observations made to M92 from June 2018 to October 2019, the M92A has timed to be an eclipsing redback with the spinning frequency of 316.5 Hz (3.16 ms) and a dispersion measure of $\sim 35.45$ pc cm$^{-3}$ [13]. The longest eclipsing event lasts for 5000 s, which corresponds to $\sim 40\%$ of the orbit. Another earliest FAST GC discovery was M13F [14], while the timing solution for all the previously known pulsars in M13 were also updated. In the systematic GC survey, ~24 new pulsars discovered from 15 GCs and timing solutions for several previously discovered pulsars were updated [10]. This made the number of GC pulsars in the FAST sky doubled (from 31 to 63). Currently, more than 40 GC pulsars were discovered by FAST[2].

---

[1] http://www.naic.edu/pfreire/GCpsr.html

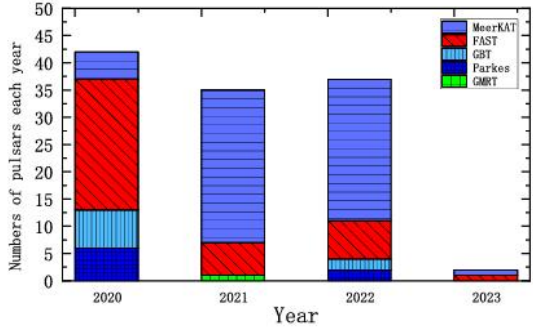

**Figure 1.** Cumulative number of new pulsars in GCs since 2020.

## 3 M15

M15 is a globular cluster with 9 pulsars currently, locating ~10 kpc away from sun [9]. Most of the pulsars in it are isolated except for M15C which has a neutron star companion and is in a 0.3-day eccentric orbit [1]. The latest discovered pulsar, M15I, is considered to be a binary [10] but no more detection till now.

Most of M15 pulsars were discovered by Arecibo [9]. The timing solution of pulsars in M15 has not been updated for years (Expect for M15A, B, and C [1]). With FAST, we have 10 observations to it (earliest is Nov. 17$^{th}$, 2018) till now and 4 long-time observations are going to come this year. All the observational data were blind searched for new pulsars, resulting in the detection of 8 known pulsars without any new discoveries. The only missing one is M15I but we have the re-detection candidate for it. Table1 shows the detection of each pulsars and our observation date. M15A, B, D, and E are always detectable, even though the observation time is only 1800 seconds. M15F and H are faint that they were only detected in long observations.

In the previous study, M15C was showing a steady decrease in brightness and significant changes in its pulse profile shape [15]. Such variations can be ascribed to relativistic spin precession. It seems that our line of sight is rapidly moving away from the magnetic axis of the visible beam. Depending on the actual size of the emitting region, M15C might be undetectable in the very near future. In out dataset, M15C was faint and still detected for 3 times (see in Figure 2). Consisted with its previous study, it seems become fainter in the date taken with FAST in 2020 than in 2018, though it is still possible to be caused by scintillation.

---
[2]https://fast.bao.ac.cn/cms/article/65/

Updated timing of pulsars in M15 still need more observations and will be reported in the future (Wu et al. in prep).

**Table 1.** M15 Pulsar Detections

| Obs | M15A | M15B | M15C | M15D | M15E | M15F | M15G | M15H | M15I |
|---|---|---|---|---|---|---|---|---|---|
| 2018.10.13 | √ | √ | √ | √ | √ | | | | |
| 2018.11.17 | √ | √ | | √ | √ | √ | | √ | |
| 2019.04.25 | √ | √ | | √ | √ | | | | |
| 2019.08.18 | √ | √ | | √ | √ | √ | | | |
| 2019.09.01 | √ | √ | | √ | √ | √ | | √ | |
| 2019.11.09 | √ | √ | | √ | √ | √ | | | |
| 2019.12.14 | √ | √ | √ | √ | √ | √ | | | |
| 2020.08.30 | √ | √ | | √ | √ | √ | | | |
| 2020.12.21 | √ | √ | | √ | √ | √ | | √ | |
| 2021.02.02 | √ | √ | | √ | √ | | √ | | |
| 2021.03.09 | √ | √ | | √ | √ | √ | | | |
| 2022.07.07 | √ | √ | √ | √ | √ | | | | |
| Total | 12 | 12 | 3 | 12 | 12 | 10 | 1 | 3 | 0 |

## 4 NGC 6517

NGC 6517 is 10.6 kpc away from Earth, being the densest cluster in the FAST coverage [16]. The first four pulsars in this GC were discovered by the Green Bank Telescope (GBT) with a average dispersion measure (DM) 182.4 pc cm$^{-3}$ [2]. NGC 6517A, C, and D are isolated while NGC 6517B is a relative young pulsar in a binary system (orbital period ~60 days, low mass companion). It was reported that FAST discovered 7 isolated pulsars NGC 6517E,F,G,H,I,K,L and a binary [10, 11]. All of them are significantly fainter than previous ones. No doubt that these discoveries are due to the high sensitivity of FAST.

Recent reprocessing of archived FAST NGC 6517 data with PulsaR Exploration and Search TOolkit (PRESTO[3]) yields the discoveries of 6 new millisecond pulsars, NGC 6517M to R[4]. The details will be described in our future paper (Yin et al. in prep). These six newly discovered pulsars are extremely faint (e.g., NGC 6517M as the example in Figure 3) and suppose to be isolated. FAST's ultra-high sensitivity can be the key to find these new pulsars.

NGC 6517 is estimated to harbour a dozen pulsars by analysing the observed flux densities of the pulsars [2]. The observed flux densities limit the mass-to-light of the cluster to prove there are no significant amounts of low-luminosity matter. From the flux limitation, it is predicted to harbor no more than 12 pulsars [2], while currently the number of pulsars is much more than this.

The high interaction rate of the core-collapsed GCs leads to a higher probability of destroying binary systems as they evolve [17]. This means isolated pulsars dominate the pulsar population in the core-collapsed GCs, and binary sys-tems are poorly populated. Pulsars in M15 and NGC 6517 are consistent with this.

---
[3]https://github.com/scottransom/presto
[4]see FAST GC survey in https://fast.bao.ac.cn/cms/article/65/, While NGC 6517R is still need to be timed.

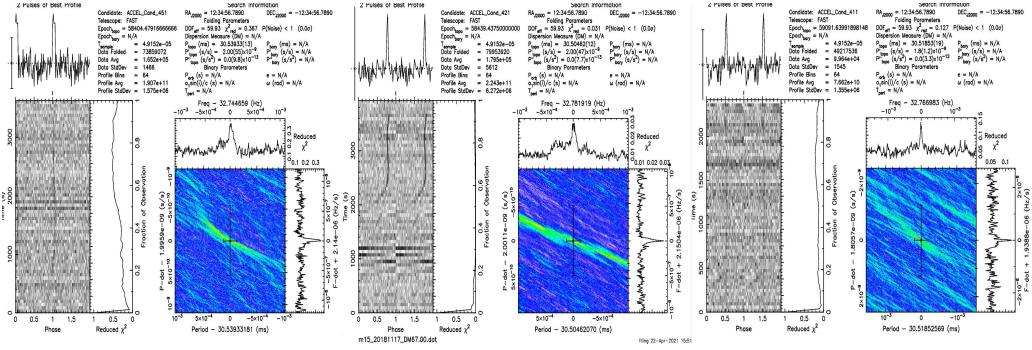

**Figure 2.** Left: M15C observed on 13$^{th}$ October 2018, Middle: M15C observed on 17$^{th}$ November 2018, Right: M15C observed on 30$^{th}$ August 2020.

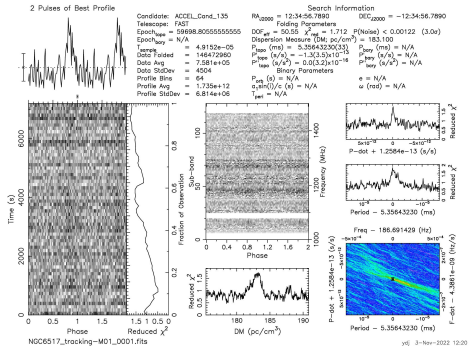

**Figure 3.** Upper: Discovery plot of NGC 6517M. Lower: initial timing residuals of NGC 6517M with JUMPs from TEMPO [18].

## 5 Conclusion

In this paper, we presented our studies on the core collapse GCs M15 and NGC 6517 with FAST. The conclusions are as follows:

1, With FAST, systematic monitoring observations to M15 and NGC 6517 started for years, resulting in more than 10 pulsars discovered from these GCs.

2, Either previously known pulsars or newly discovered, most of pulsars in these two GCs are isolated. This is consisted with the fact that the two body encounter rate are high in core collapse GCs thus the binary systems are highly possibly to be destroyed.

3, M15C, which has a neutron star companion, was still detectable by FAST. It became fainter from 2018 to 2020, may due to the precession and/or scintillation.

4, Newly discovered pulsars in NGC 6517 are all tend to be isolated. The number of pulsars in NGC 6517 now is much more than the prediction by Lynch et al [2].

## 6 Acknowledgement

This work is supported by SKA Program of China No. 2020SKA0120200, National Key R&D Program of China No. 2021YFA1600403. and National Science Foundation, China, No. 11703047, 11773041, U1931128, U2031119, 12173052, 12003047, 12173053, 1204130, 12073071, 11988101, 11833009, 11873058, U1731120, No. 11963002. Z. P. and L. Q. are or were supported CAS "Light of West China" Program. L. Q. was supported by Youth Innovation Promotion Association of CAS (ID 2018075). We also thank the fostering project of GuiZhou University with No. 201911, and Cultivation Project for FAST Scientific Payoff and Research Achievement of CAMS-CAS. This work made use of data from the Five-hundred-meter Aperture Spherical radio Telescope (FAST). FAST is a Chinese national mega-science facility, built and operated by the National Astronomical Observatories, Chinese Academy of Sciences (NAOC). We appreciate all the people from the FAST group for their support and assistance during the observations.

## References

[1] Jacoby B A, Cameron P B, Jenet F A, et al, "Measurement of orbital decay in the double neutron star binary PSR B2127+ 11C," *The Astrophysical Journal*, 2006, 644(2): L113. doi: 10.1086/505742.

[2] Lynch, R. S., Ransom, S. M., Freire, P. C. C., and Stairs, I. H., "Six New Recycled Globular Cluster Pulsars Discovered with the Green Bank Telescope," *The Astrophysical Journal*, vol. 734, no. 2, 2011. doi:10.1088/0004-637X/734/2/89.

[3] Hewish A, Bell S J, Pilkington J D H, et al, "74. observation of a rapidly pulsating radio source," *A Source Book in Astronomy and Astrophysics*, 1900–1975, Harvard University Press, 2013, 498-504.

[4] Nan R, Li D, Jin C, et al, "The five-hundred-meter aperture spherical radio telescope (FAST) project," *International Journal of Modern Physics D*, 20.06 (2011): 989-1024. doi: 10.1109/mwp.2015.7356696.

[5] Li D, Wang P, Qian L, et al, " FAST in space: considerations for a multibeam, multipurpose survey using China's 500-m aperture spherical radio telescope (FAST)," *IEEE Microwave Magazine*, 2018, 19(3): 112-119. doi:10.1109/mmm.2018.2802178

[6] Han J L, Wang C, Wang P F, et al, "The FAST Galactic Plane Pulsar Snapshot survey: I. Project design and pulsar discoveries," *Research in Astronomy and Astrophysics*, 2021, 21(5): 107. doi:10.1088/1674-4527/21/5/0n1 .

[7] Hessels J W T, Ransom S M, Stairs I H, et al, "A radio pulsar spinning at 716 Hz," *Science*, 20.06 (2011):311.5769 (2006): 1901-1904. doi: 10.1126/science.1123430.

[8] Prince T A, Anderson S B, Kulkarni S R, et al, "Timing Observations of the 8-Hr Binary Pulsar 2127+ 11C in the Globular Cluster M15," *Astrophysical Journal*, 374 (1991): L41-L44. doi: 10.1086/186067.

[9] Anderson, Stuart Bruce, A study of recycled pulsars in globular clusters, *California Institute of Technology*, 1993.

[10] Pan Z, Qian L, Ma X, et al, "FAST Globular Cluster Pulsar survey: twenty-four pulsars discovered in 15 globular clusters," *The Astrophysical Journal Letters*, 915.2 (2021): L28. doi: 10.3847/2041-8213/ac0bbd.

[11] Pan, Z., "Three pulsars discovered by FAST in the globular cluster NGC 6517 with a pulsar candidate sifting code based on dispersion measure to signal-to-noise ratio plot," *Research in Astronomy and Astrophysics*, vol. 21, no. 6, 2021. doi:10.1088/1674-4527/21/6/143.

[12] Qian L, Pan Z C, Li D, et al, "The first pulsar discovered by FAST," *Science China Physics, Mechanics & Astronomy*, 2019, 62: 1-4. doi: 10.1007/s11433-018-9354-y.

[13] Pan Z, Ransom S M, Lorimer D R, et al, "The FAST discovery of an eclipsing binary millisecond pulsar in the globular cluster M92 (NGC 6341)," *The Astrophysical Journal Letters*, 2020, 892(1): L6. doi: 10.3847/2041-8213/ab799d.

[14] Wang L, Peng B, Stappers B W, et al, "Discovery and timing of pulsars in the globular cluster M13 with FAST," *The Astrophysical Journal*, 2020, 892(1): 43. doi: 10.1007/978-94-011-2704-2_15.

[15] Ridolfi A, Freire P C C, Kramer M, et al, " Long-term observations of pulsars in the globular clusters 47 Tucanae and M15," *Proceedings of the International Astronomical Union*, 2017, 13(S337): 251-254. doi: 10.1017/s1743921317009334.

[16] Harris, W. E., "A Catalog of Parameters for Globular Clusters in the Milky Way," *The Astronomical Journal*, vol. 112, p. 1487, 1996. doi:10.1086/118116.

[17] Verbunt, F. and Freire, P. C. C., "On the disruption of pulsar and X-ray binar ies in globular clusters", *Astronomy and Astrophysics*, vol. 561, 2014. doi:10.1051/0004-6361/201321177.

[18] Nice, D., "Tempo: Pulsar timing data analysis", *Astrophysics Source Code Library*, 2015. ascl:1509.002.